\documentclass[manuscript]{aastex}

\shorttitle{A close substellar companion to HD\,149382}
\shortauthors{Geier et al.}

\begin{document}


\title{Discovery of a close substellar companion to the hot subdwarf star HD\,149382\\
    -- The decisive influence of substellar objects on late stellar evolution}


\author{S. Geier, H. Edelmann\altaffilmark{1} and U. Heber}
\affil{Dr.\,Remeis-Sternwarte, Institute for
 Astronomy, University Erlangen-N\"urnberg, Sternwartstr.~7, 96049
 Bamberg, Germany}
\email{geier@sternwarte.uni-erlangen.de}

\and

\author{L. Morales-Rueda}
\affil{Department of Astrophysics, Faculty of Science, Radboud University Nijmegen, P.O. Box 9010, 6500 GL Nijmegen, NE}


\altaffiltext{1}{McDonald Observatory, University of Texas at Austin, 1 University Station, C1402, Austin, TX 78712-0259, USA}


\begin{abstract}

Substellar objects, like planets and brown dwarfs orbiting stars, are by-products of the star formation process. The evolution of their host stars may have an enourmous impact on these small companions. Vice versa a planet might also influence stellar evolution as has recently been argued. 

Here we report the discovery of a 8$-$23 Jupiter-mass substellar object orbiting the hot subdwarf HD\,149382 in 2.391 days at a distance of only about five solar radii. Obviously the companion must have survived engulfment in the red-giant envelope. Moreover, the substellar companion has triggered envelope ejection and enabled the sdB star to form. Hot subdwarf stars have been identified as the sources of the unexpected ultravoilet emission in elliptical galaxies, but the formation of these stars is not fully understood. Being the brightest star of its class, HD~149382 offers the best conditions to detect the substellar companion. Hence, undisclosed substellar companions offer a natural solution for the long-standing formation problem of apparently single hot subdwarf stars. Planets and brown dwarfs may therefore alter the evolution of old stellar populations and may also significantly affect the UV-emission of elliptical galaxies.

\end{abstract}





\section{Introduction}

A long-standing problem in extragalactic astronomy is the UV-excess observed in the spectra of elliptical galaxies. This phenomenon is caused by an old population of helium burning stars, known as hot subdwarfs or sdBs \citep[see review by][]{heber}. The origin of the UV-excess can, hence, be traced back to that of the sdB stars themselves. The formation of such stars remains a mystery as it requires an extraordinarily high mass loss on the red-giant branch. Hot subdwarfs often reside in close binaries, formed by ejection of the envelope of their red-giant progenitors through interaction with the stellar companion. However, for half of the known sdBs no such companions could be found, requiring a yet unknown sdB formation channel.

After finishing core-hydrogen-burning the progenitors of sdBs leave the main sequence and evolve to red giants before igniting helium and settling down on the extreme horizontal branch. Unlike normal stars, the sdB progenitors must have experienced a phase of extensive mass loss on the red giant branch, in order to explain the high temperatures and gravities observed at the surface of hot subdwarf stars. After consumption of the helium fuel they evolve directly to white dwarfs avoiding a second red-giant phase. What causes this extensive mass loss remains an open question.  

The riddle of sdB formation is closely related to other long-standing problems regarding old stellar populations, which have been discussed for decades. The morphology of the horizontal branch in globular clusters, especially the existence and shape of its extreme hot part (Extreme Horizontal Branch, EHB), is still far from understood \citep{catelan}. Hot subdwarfs are also regarded as the dominant source of the UV-excess in early type galaxies, where no active star formation is going on and hence no UV-emission from young massive  stars is expected . Hot subdwarf formation is the key to understanding the physics behind this phenomenon and a debate is going on whether single star \citep{yi} or binary evolution \citep{han3} explain the observed UV-excesses.

About half of the sdB stars reside in close binaries with periods as short
as $\sim 0.1$\,d \citep{maxted1,napiwotzki1}. Because the components' separation in these systems is much less than the size of the subdwarf progenitor in its red-giant
phase, these systems must have experienced a common-envelope and
spiral-in phase \citep{han1,han2}. In such a scenario, two
main-sequence stars of different masses evolve in a binary system. The more massive one will first reach the red-giant phase and at some point
fill its Roche lobe, where mass is transferred from the giant to the
companion star. When mass transfer is unstable, the envelope of the
giant will engulf the companion star and form a common envelope. Due
to friction with the envelope, the two stellar cores lose orbital
energy and spiral towards each other until enough orbital energy has
been deposited within the envelope to eject it. The end
product is a much closer system containing the core of the giant,
which then may become an sdB star, and a main-sequence companion. 
This companion evolves to a white dwarf after another phase of unstable mass 
transfer. 

The common-envelope ejection channel provides a reasonable explanation for the extra mass loss required to form sdB stars. But for about half of all analysed subdwarfs there is no evidence for close stellar companions as no radial velocity variations are found. Although in some cases main sequence companions are visible in the spectra, it remains unclear whether these stars are close enough to have interacted with the sdB progenitors. Among other formation scenarios, the merger of two helium white dwarfs has often been suggested to explain the origin of single sdB stars \citep{han1,han2}. Merging should result in rapidly spinning stars, which is not consistent with observations. A recent analysis of single sdB stars revealed that their $v_{\rm rot}\sin{i}$-distribution is consistent with a uniform rotational velocity $v_{\rm rot}\approx8\,{\rm km~s^{-1}}$ and randomly oriented polar axes \citep{geier3}.

The planet discovered to orbit the sdB pulsator V\,931\,Peg with a period of $1\,170\,{\rm d}$ and a separation of $1.7\,{\rm AU}$ was the first planet found to have survived the red-giant phase of its host star \citep{silvotti}. Serendipitous discoveries of two substellar companions around the eclipsing sdB binary HW\,Vir at distances of $3.6\,{\rm AU}$ and $5.3\,{\rm AU}$ with orbital periods of $3\,321\,{\rm d}$ and $5\,767\,{\rm d}$ \citep{lee} and one brown dwarf around the similar system HS\,0705$+$6700 with a period of $2\,610\,{\rm d}$ and a separation of less than $3.6\,{\rm AU}$ \citep{qian} followed recently. These substellar companions to hot subdwarfs have rather wide orbits, were not engulfed by the red giant progenitor and therefore could not have influenced the evolution of their host stars.

\section{Observations and Analysis}

HD\,149382 is the brightest core helium-burning subdwarf known. The first hint that this star could show very small radial velocity (RV) variations was found during our survey aimed at finding sdBs in long period binaries \citep{edelmann}. We obtained 15 high resolution spectra ($R=30\,000-48\,000$) within four years with three different high resolution spectrographs (ESO-2.2m/FEROS, McDonald-2.7m/Coud\'e, CAHA-2.2m/FOCES). One additional spectrum obtained with ESO-VLT/UVES at highest resolution ($R\approx80\,000$) was taken from the ESO-archive. 

In order to measure the RVs with highest possible accuracy we fitted a set of mathematical functions (polynomial, Gaussian \& Lorentzian) to all suitable spectral lines with wavelengths from about $4\,000\,{\rm \AA}$ to $6\,700\,{\rm \AA}$ using the FITSB2 routine \citep{napiwotzki2}. The formal deviation along the whole wavelength range was $0.2\,{\rm km~s^{-1}}$ at best. In order to check the accuracy of this measurements, we also obtained RVs from telluric and night sky lines and reached similar accuracies. Since telluric and night sky lines have zero RV we used them to correct the measured RVs for calibration errors. The applied corrections were usually below $1.0\,{\rm km~s^{-1}}$. Since we used four entirely different instruments and obtained consistent results other systematic effects should be neligible. The RV measurements are given in Table~\ref{RVs}.

\begin{figure}
\includegraphics[width=16cm]{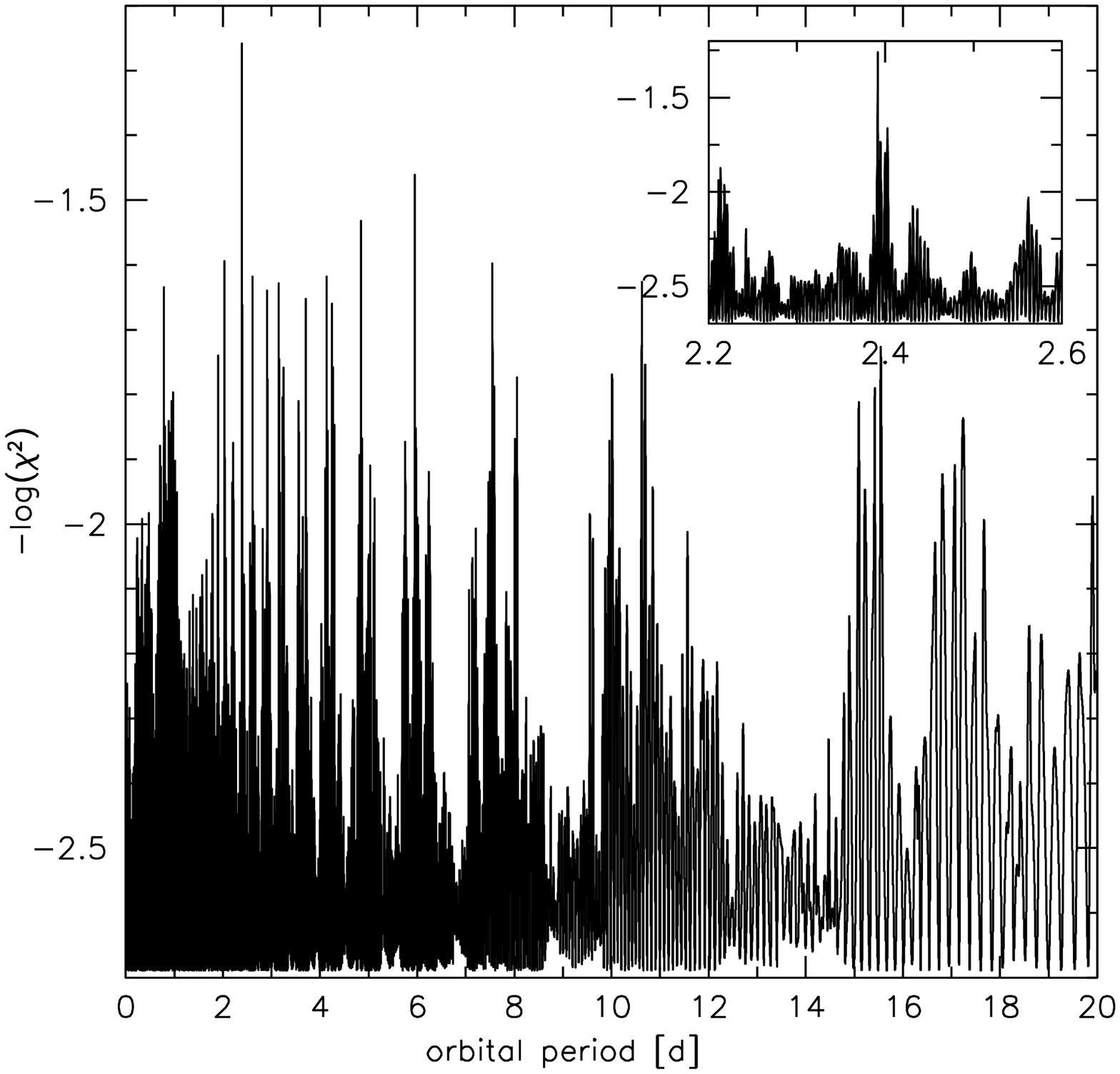}
 \caption{Power spectrum of HD\,149382. $-\log{\chi^{2}}$ is plotted against the orbital period in days. The region around the most prominent period is plotted in the small window. The formal significance exceeds the $3\sigma$-limit.}
 \label{power}
\end{figure}

\begin{figure}
 \includegraphics[width=16cm]{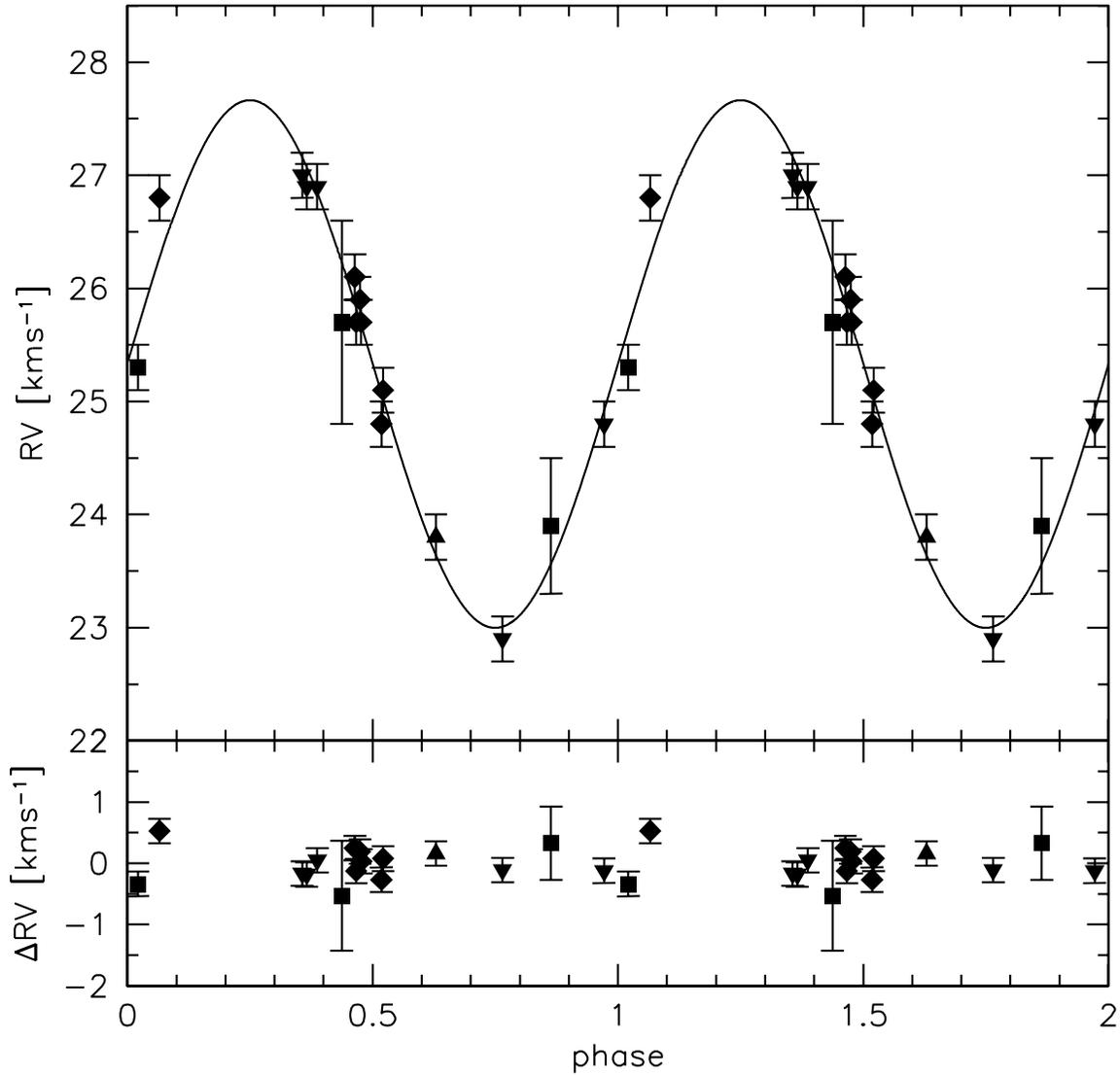}
 \caption{Radial velocity curve of HD\,149382. The plot shows the radial velocity plotted against orbital phase (diamonds: McDonald-2.7m/Coud\'e, rectangles: CAHA-2.2m/FOCES, upside down triangles: ESO-2.2m/FEROS, triangle: ESO-VLT/UVES). The RV data was folded with the most likely orbital period. The residuals are plotted below.}
 \label{rv}
\end{figure}

The period search was carried out by means of a periodogram based on the Singular Value Decomposition (SVD) method. A sine-shaped RV curve was fitted to the observations for a multitude of phases, which were calculated as a function of period. The difference between the observed radial velocities and the best fitting theoretical RV curve for each phase set was evaluated in terms of the logarithm of the sum of the squared residuals ($\chi^{2}$) as a function of period. This method finally results in the power spectrum of the data set which allows to determine the most probable period $P$ of variability \citep{lorenz}. The formal significance of the best orbital solution ($P=2.391\pm0.002\,{\rm d}$, $K=2.3\pm0.1\,{\rm km~s^{-1}}$, $\gamma=25.3\pm0.06\,{\rm km~s^{-1}}$) exceeds the $3\sigma$-limit (see Figure~\ref{power}) and a very small mass function $f(M)=3.8\times10^{-6}$ results. The radial velocity curve is shown in Fig.~\ref{rv}. The formal probability that the next best alias periods at about $4.8\,{\rm d}$ and $6.0\,{\rm d}$ are correct is less than $5\,\%$. Even if one of these longer periods should be the correct one, the mass function increases only by a factor of $1.5$ at most and our conclusions still remain valid. 

The atmospheric parameters effective temperature $T_{\rm eff}$, surface gravity $\log\,g$ and helium abundance were determined by fitting simultaneously 17 hydrogen and helium lines in high resolution, high-S/N FEROS and UVES spectra with NLTE model spectra \citep[the method is described in ][]{lisker,geier1}. The parameters ($T_{\rm eff}=35\,500\pm500\,{\rm K}$, $\log\,g=5.80\pm0.05$) are in good agreement with the result of \citet{saffer}: $T_{\rm eff}=34\,200\pm1\,500\,{\rm K}$, $\log\,g=5.89\pm0.15$. 

The mass of the unseen companion can be derived by solving the binary mass function $f_{\rm m} = M_{\rm comp}^3\sin^3i/(M_{\rm comp} + M_{\rm sdB})^2 = P K^3/2 \pi G$. In order to obtain a unique solution, the mass of the sdB primary as well as the inclination of the system must be known. Due to the excellent quality of the data available for HD\,149382, constraints can be put on both crucial parameters. 

The distance to this star can be derived directly using a trigonometric parallax obtained with the HIPPARCOS satellite \citep{leeuwen}. We derive the angular diameter by comparing the surface flux in the V band computed from a model atmosphere with the derived atmospheric parameters to the observed value \citep{mermilliod}. Using the trigonometric distance we can derive the stellar radius, and from the surface gravity the mass of the sdB \citep{ramspeck}. 

Taking the uncertainties of all parameters into account (V magnitude, $T_{\rm eff}$, $\log{g}$, parallax) the possible mass range for the sdB is $0.29-0.53\,M_{\rm \odot}$. This range is consistent with the canonical mass of $0.47\,M_{\rm \odot}$ derived from single and binary evolution calculations \citep{han1,han2}. Without further constraints on the inclination only a lower limit to the mass of the unseen companion can be calculated. 

The minimum companion mass lies between $0.006\,M_{\rm \odot}$ and $0.01\,M_{\rm \odot}$, well below the stellar limit of $0.075\,M_{\rm \odot}$ to $0.083\,M_{\rm \odot}$ depending on the metallicity \citep{chabrier}, which is the lower limit where core hydrogen burning can be ignited and a star can be formed. The lower the inclination of the binary is, the higher is the mass of the unseen companion. Assuming a random distribution of orbital plane inclinations, the probability for a binary to fall below a certain inclination can be derived \citep{gray}. The probability for the companion to have a mass of more than $0.08\,M_{\rm \odot}$ is just $0.8\,\%$. The probability that the mass of the unseen companion exceeds the planetary limit of $0.012\,M_{\rm \odot}$ defined by the IAU\footnote[1]{Position statement on the definition of a "planet". Working group on extrasolar planets of the International Astronomical Union. http://www.dtm.ciw.edu/boss/definition.html} is only $33\,\%$. 
 
However, we can constrain the mass of HD\,149382\,b even further. Due to the very high resolution of the UVES spectrum the broadening caused by the projected rotational velocity of the star could be measured from the metal lines although it turned out to be as small as $v_{\rm rot}\sin{i}=4.9\pm1.4\,{\rm km~s^{-1}}$. In order to derive $v_{\rm rot}\,\sin{i}$, we compared the observed high resolution ($R=80\,000$) UVES spectrum with rotationally broadened, synthetic line profiles. The profiles were computed using the LINFOR program \citep{lemke}. A simultaneous fit of elemental abundance, projected rotational velocity and radial velocity was then performed separately for every identified line using the FITSB2 routine \citep{napiwotzki2}. The method is described in \citet{geier2}.

Assuming that HD\,149382 rotates with the standard velocity of $8\,{\rm km~s^{-1}}$ infered for single sdBs \citep{geier3} the inclination can be constrained to $i=26^{\circ}-52^{\circ}$. The companion mass range is derived to be $M_{\rm 2}=0.008-0.022\,M_{\rm \odot}=8-23\,M_{\rm J}$ consistent with a gas giant planet or a low mass brown dwarf. Adopting the statistically most likely inclination $i=52^{\circ}$ and the canonical sdB mass of $0.47\,M_{\rm \odot}$ \citep{han1,han2} the companion mass is $0.011\,M_{\rm \odot}=12\,M_{\rm J}$, which places HD\,149382\,b just below the planetary limit. The separation of star and companion is $5-6\,R_{\rm \odot}$. All relevant measurements and parameters of the HD\,149382 system are summarized in Table~\ref{params}. 

\section{Discussion}

When the progenitor of HD\,149382 evolved through the red-giant phase, it expanded to a radius of ten times the present orbital separation. The initial separation must have been larger (about 1 AU) and the companion spiralled-in due to interaction with the giant's envelope until the envelope was ejected. Despite the very high local temperature inside the envelope \citep[$300\,000$ to $400\,000\,{\rm K}$ at $5-6\,R_{\rm \odot}$ from the giant's centre,][]{soker} the substellar companion survived. The companion of the sdB star AA~Dor has also been suggested to be a brown dwarf in a $0.26\,{\rm d}$ orbit \citep{rauch,rucinski}. This conclusion, however, is rendered uncertain as \citet{vuckovic} derive a higher mass indicating that the companion is a star.

\citet{soker} suggested that sub-stellar objects like brown dwarfs and planets may also be swallowed by their host star and that common-envelope ejection could form hot subdwarfs. Substellar objects with masses higher than $\approx10\,M_{\rm J}$ were predicted to survive the common-envelope phase and end up in a close orbit around the stellar remnant, while planets with lower masses would entirely evaporate. The stellar remnant is predicted to lose most of its envelope and evolve towards the EHB. The orbital period and mass we derived for HD\,149382\,b is in excellent agreement with the predictions made by \citet{soker}. A similar scenario has been proposed to explain the formation of apparently single low mass white dwarfs \citep{nelemans}. The discovery of a brown dwarf with a mass of $0.053\pm0.006\,M_{\rm \odot}$ in a $0.08\,{\rm d}$ orbit around such a white dwarf supports this scenario and shows that substellar companions can influence the outcome of stellar evolution \citep{maxted2}.

The discovery of planets and brown dwarfs around sdBs and especially the close-in substellar companion of HD\,149382 may thus have important implications for the still open question of sdB formation. The extraordinary quality of the photometric data was a prerequisite for the detection of the substellar companions in V\,931\,Peg, HW\,Vir and HS\,0705$+$6700 \citep{silvotti,lee,qian}. Finding such companions orbiting three of the best observed sdBs cannot be mere conincidence and leads to the conclusion that substellar objects may often be associated with sdBs. HD\,149382 is the brightest sdB known. Hence the quality of the spectroscopic data is also very high. It is not easy to detect such small RV variations even in high resolution spectra. The fact that we found them in the case of HD\,149382 leads to the conclusion that close-in planets or brown dwarfs may be common around apparently single sdB stars. They were just not detected up to now. Hence all apparently single sdBs may have or had close brown dwarf or planetary companions, although those of lowest mass may have evaporated.

HD\,149382\,b provides evidence that substellar companions can decisively change the evolution of stars, as they trigger extensive mass loss. They could be responsible for the formation of the single hot subdwarf population. These stars are not only numerous in our Galaxy, but also make elliptical galaxies shine in ultraviolet light.




\acknowledgments

This letter was based on observations at the La Silla Observatory of the European Southern Observatory for programme number 077.D-0515(A). Some of the data used in this work were obtained at the McDonald Observatory. Based on observations collected at the Centro Astron\'omico Hispano Alem\'an (CAHA) at Calar Alto, operated jointly by the Max-Planck Institut f\"ur Astronomie and the Instituto de Astrof\'isica de Andaluc\'ia (CSIC). Some of the data used in this work were downloaded from the ESO archive. S.G. gratefully acknowledges financial support from the Deutsche Forschungsgemeinschaft through grant He~1356/40-4. We thank Sebastian M\"uller for reducing the archival UVES spectrum.

\clearpage



\clearpage

\clearpage

\clearpage

\begin{table}
\caption{\bf Radial velocities of HD\,149382}
\label{RVs}
\begin{center}
\begin{tabular}{lll}
\tableline
\noalign{\smallskip}
mid-HJD & RV [${\rm km~s^{-1}}$] & Instrument\\
\noalign{\smallskip}
\tableline
\noalign{\smallskip}
2452497.49150 &  $27.0\pm0.2$ & FEROS \\
2452497.51327 &  $26.9\pm0.2$ & FEROS \\
2452497.56524 &  $26.9\pm0.2$ & FEROS \\
2452891.30690 &  $25.3\pm0.2$ & FOCES \\
2452892.30170 &  $25.7\pm0.9$ & FOCES \\ 
2452893.32160 &  $23.9\pm0.6$ & FOCES \\
2453784.83294 &  $23.8\pm0.2$ & UVES \\ 
2453904.73904 &  $22.9\pm0.2$ & FEROS \\ 
2453931.76614 &  $26.8\pm0.2$ & Coud\'e \\
2453932.71806 &  $26.1\pm0.2$ & Coud\'e \\
2453932.72485 &  $25.7\pm0.2$ & Coud\'e \\
2453932.74337 &  $25.9\pm0.2$ & Coud\'e \\
2453932.74882 &  $25.7\pm0.2$ & Coud\'e \\
2453932.84790 &  $24.8\pm0.2$ & Coud\'e \\
2453932.85560 &  $25.1\pm0.2$ & Coud\'e \\
2453986.54830 &  $24.8\pm0.2$ & FEROS \\
\noalign{\smallskip}
\tableline

\end{tabular}
\end{center}
\end{table}

\clearpage

\begin{table}
\caption{\bf Parameters of the HD\,149382 system}
\label{params}
\begin{center}
\begin{tabular}{llll}

\tableline
\noalign{\smallskip}

Trigonometric parallax\tablenotemark{a} & $\pi$ & [mas] & $13.53\pm1.15$\\
Distance & $d$ & [pc] & $74^{+7}_{-8}$\\
Visual magnitude\tablenotemark{b} & $m_{\rm V}$ & [mag] & $8.947\pm0.009$\\

\noalign{\smallskip}
\tableline
\noalign{\smallskip}

Atmospheric parameters & & & \\

\noalign{\smallskip}
\tableline
\noalign{\smallskip}

Effective temperature & $T_{\rm eff}$ & [K] & $35\,500\pm500$\\
Surface gravity & $\log{g}$           & & $5.80\pm0.05$\\
Helium abundance & $\log{y}$          & & $-1.44\pm0.01$\\
Projected rotational velocity & $v_{\rm rot}\sin{i}$ & [${\rm km~s^{-1}}$] & $4.9\pm1.4$\\

\noalign{\smallskip}
\tableline
\noalign{\smallskip}

Orbital parameters & & & \\

\noalign{\smallskip}
\tableline
\noalign{\smallskip}

Orbital period & $P$ & [d] & $2.391\pm0.002$\\
RV semi-amplitude & $K$ & [${\rm km~s^{-1}}$] & $2.3\pm0.1$\\
System velocity & $\gamma$ & [${\rm km~s^{-1}}$] & $25.3\pm0.06$\\
Binary mass function & $f(M)$ & [$M_{\rm \odot}$] & $3.8\times10^{-6}$\\

\noalign{\smallskip}
\tableline
\noalign{\smallskip}

Derived parameters & & & \\

\noalign{\smallskip}
\tableline
\noalign{\smallskip}

Subdwarf mass & $M_{\rm sdB}$ & [$M_{\rm \odot}$] & $0.29-0.53$\\
Orbital inclination & $i$ & [$^{\rm \circ}$] & $26-52$ \\
Companion mass & $M_{\rm comp}$ & [$M_{\rm J}$] & $8-23$\\
                         &                & [$M_{\rm \odot}$] & $0.008-0.022$\\
Separation & $a$ & [$R_{\rm \odot}$] & $5.0-6.1$\\

\tableline\\
\end{tabular}
\end{center}
\tablenotetext{a}{The trigonometric parallax was taken from the new reduction of the HIPPARCOS data \citep{leeuwen}.}
\tablenotetext{b}{The visual magnitude is taken from \citet{mermilliod}.} 
\end{table}




\end{document}